# Deep Learning for Classification of Thyroid Nodules on Ultrasound: Validation on an Independent Dataset

Jingxi Weng[1], Benjamin Wildman-Tobriner[2], Mateusz Buda[3], Jichen Yang[3], Lisa M. Ho[4], Brian C. Allen[4], Wendy L. Ehieli[4], Chad M. Miller[4], Jikai Zhang[3] and Maciej A. Mazurowski[2]

1: Department of Radiation Oncology, University of Florida, Gainesville, FL, USA
2: Department of Radiology, Duke University, Durham, NC, USA
3: Department of Electrical and Computer Engineering, Duke University, Durham, NC, USA
4: Department of Radiology, Duke University Medical Center

Correspondence Author:
Jichen Yang,
jy168@duke.edu,
Duke University, Department of Electrical and Computer Engineering
1809 Wrenn Rd
Durham. NC, USA 27708-0187
984-244-9678

**Abstract**

**Objectives:** The purpose is to apply a previously validated deep learning algorithm to a new thyroid nodule ultrasound image dataset and compare its performances with radiologists.

**Methods:** Prior study presented an algorithm which is able to detect thyroid nodules and then make malignancy classifications with two ultrasound images. A multi-task deep convolutional neural network was trained from 1278 nodules and originally tested with 99 separate nodules. The results were comparable with that of radiologists. The algorithm was further tested with 378 nodules imaged with ultrasound machines from different manufacturers and product types than the training cases. Four experienced radiologists were requested to evaluate the nodules for comparison with deep learning.

**Results:** The Area Under Curve (AUC) of the deep learning algorithm and four radiologists were calculated with parametric, binormal estimation. For the deep learning algorithm, the AUC was 0.69 (95% CI: 0.64 – 0.75). The AUC of radiologists were 0.63 (95% CI: 0.59 – 0.67), 0.66 (95% CI:0.61 - 0.71), 0.65 (95% CI: 0.60 – 0.70), and 0.63 (95%CI: 0.58 – 0.67).

**Conclusion:** In the new testing dataset, the deep learning algorithm achieved similar performances with all four radiologists. The relative performance difference between the algorithm and the radiologists is not significantly affected by the difference of ultrasound scanner.

**Introduction**

Deep learning has been increasingly applied to tasks involving medical imaging [1] and in some instances has demonstrated performance similar to or better than radiologists [2]. While promising results are often exciting and may illuminate potential advantages of integrating deep learning with diagnostic imaging, many studies share common and important problems. One issue common to many deep learning studies is that training and validation data are often derived from one pool of cases from a single institution. Developing and testing a model from a single source can be limiting and may lead to poor generalizability and low performance in new environments. Therefore, it is crucial to test deep learning models with data from other institutions, which use scanners and protocols different from those used during model development.

With these factors in mind, we aimed to evaluate the performance of a deep learning algorithm designed for thyroid nodule classification. Several studies have shown that deep learning can perform well for the task of interpreting thyroid nodules [3–6]. However, our model was developed using cases from a single institution, and our goal was to test its performance on a larger, more robust set of cases from a different institution.

The purpose of our study was to take a deep learning neural network that previously showed radiologist-level performance in evaluating thyroid nodules and apply it to fully independent data from a different institution. Our goal was to evaluate how an independent dataset might change overall model performance compared to human readers and to assess the impact of scanner types on performance.

**Material and Methods**

This retrospective study was approved by our institutional review board and was HIPAA compliant. The informed consent from patients is waived by IRB protocols.

*Study Population*

378 thyroid nodules from 320 patients were included in the study. A single transverse and longitudinal image of each nodule was included for a total of 756 ultrasound images. This dataset was collated from a larger dataset containing 3683 patients from electronic medical records at our institution. First, 668 patients without available ultrasound images within 90 days prior to the pathology report date were excluded. Then, 421 patients were excluded due to follicular nodules. Due to the busy schedule of radiologists, the radiologists were only able to review 426 nodules out of the whole dataset. This subset was not selected in any particular way, and was only selected based on the numerical order of filename. And then, identical images, mismatched images, single-view images, images of non-thyroid nodules, and images with extra or deficient calipers were excluded, which made the final dataset 320 patients with 378 nodules, as shown in Figure 1. In the final dataset, all nodules had two corresponding images from two directions with 4 calipers on one image, and 2 or 4 calipers on another image. All of the exclusion and selection of the dataset were examined and checked by experienced radiologists.

Images were acquired from 9 different ultrasound device types during routine clinical scanning: GE LOGIQ E9 (79/378, 20.90%) (GE, Boston, MA, USA), GE LOGIQ 9

(1/378, 0.26%) (GE, Boston, MA, USA), Philips ATL HDI 3000 (4/378, 1.06%) (Philips, Amsterdam, Netherlands), Philips ATL HDI 5000 (64/378, 16.93%) (Philips, Amsterdam, Netherlands), Philips iU22 (218/378, 57.67%) (Philips, Amsterdam, Netherlands), Siemens ACUSON Sequoia (1/378, 0.26%) (Siemens, Munich, Germany), Siemens S2000 (5/378, 1.31%) (Siemens, Munich, Germany), ZONARE Z_ONE (1/378, 0.26%) (Mindray North America, Mahwah, NJ, USA), and a small group of images were processed with MPTronic software (5/378, 1.32%) (MPTronic, Paris, France). Since the appearance of images processed with MPTronic software looked different due to post-processing, these images were treated as a separate category for analysis. Images from Philips ATL HDI 5000 (Philips, Amsterdam, Netherlands), GE LOGIQ E9 (GE, Boston, MA, USA), and Philips iU22 (Philips, Amsterdam, Netherlands) in total consisted of 95.50% of all images, and we chose to focus on these three main device types.

*Pathological Ground Truth*

All 378 nodules had undergone fine needle aspiration (FNA), and FNA results according to the Bethesda criteria were considered ground truth [7]. The benign nodules included were all Bethesda 2, while nodules considered malignant were either Bethesda 5 (suspicious for malignancy) or 6 (diagnostic of malignancy). Bethesda 3 or 4 were not included in the study.

Out of the 378 nodules (756 US images) analyzed in this study, 147 were malignant and 231 were benign. Nodule and patient characteristics are listed in Table 1.

*Reader study*

Four radiologist readers with 2, 9, 15, and 22 years of post-fellowship experience participated. Each radiologist individually reviewed every nodule and subsequently answered two questions for each nodule: 1) Would you recommend FNA of this nodule (Yes/No) and 2) What is your intuition as to the likelihood of malignancy (Almost certainly benign, Probably benign, Moderate chance of malignancy, Probably malignant, and Almost certainly malignant). Readers were instructed to use whatever clinical practice or guidelines they routinely used for thyroid nodule evaluation prior to the introduction of the American College of Radiology Thyroid Imaging Reporting and Data System (ACR TI-RADS) [8], as at our institution at the time of this study, ACR TI-RADS was only just being introduced and was inconsistently used. The reader study was finished in Fall 2020, but the study started when TI-RADS was just introduced. Most readers used a combination of the American Thyroid Association (ATA) guidelines and their own clinical acumen. In addition to these instructions, readers were told there may be more malignant nodules than are typically seen in routine practice, though they were not given a specific frequency with which to expect malignant nodules.

In order to examine the agreement between the responses of radiologists, we have applied Cohen's kappa analysis [9]. Apparently, the radiologists had different understandings of likelihood of malignancy. Since TI-RADS was not used in the study to decide the likelihood of malignancy and the judgment of likelihood could be affected by personal understandings, the response "Moderate chance of malignancy" and "Probably benign" were combined together for kappa analysis. In that case, the discrepancy caused by different understandings of radiologists as well as not using TI-RADS could be reduced.

The responses of the second question were converted into numerical integers from 1 (Almost certainly benign) - 5 (Almost certainly malignant) for analyses. We have also calculated the response score based on the average scores of four radiologists, which is referred as radiologist average.

*Deep Learning Algorithm*

Our team previously developed a deep learning algorithm for risk stratification of thyroid nodules on ultrasound [6]. Briefly, the model was based on 1631 nodules from another institution and were acquired from several manufacturers: 94.4% from Siemens Antares and Elegra, 4.0% from GE LOGIQ E9, 0.9% from Philips iU22, and 0.4% from Philips ATL HDI 5000. No nodules from our own institution were used in initial model development. Initial model testing was performed using 99 nodules, 15 of which were malignant. In the previous paper, AUC value of radiologist average and deep learning are 0.80 and 0.78 in the training dataset, 0.82 and 0.87 in the test dataset.

While the exact details of the algorithm have been described, a brief review is warranted. The model consists of three parts: nodule detection, nodule classification, and risk level stratification.

Nodules are detected based on the caliper placement in each image. To detect the calipers, we trained an image detection network based on Faster Region-based Convolutional Neural Network (Faster R-CNN) with ResNet-101 backbone pre-trained on MS COCO dataset [10–12]. After detection, a square with 32-pixel margin around calipers was selected from the original image and resized into 160x160 pixels.

For classification, a deep convolutional neural network was trained from scratch, which can be used to predict the probability of benignity or malignancy of one nodule. The architecture of the network is shown in Figure 2. It consisted of six $3 \times 3$ convolutional filters, with ReLU activation function and five $2 \times 2$ max pooling layers. Then, 50% dropout [13] was added for regularization in the training phase. Finally, one fully connected layer with one output is assigned and sigmoid function is used. The final output of the network was the probability of malignancy.

*Statistical analysis of performance*

We evaluated the performance of radiologists and deep learning algorithm for distinguishing benign from malignant nodules.

The performance of radiologists was calculated based on their assignment of the likelihood of malignancy which were encoded as ordinal numbers. The performance of deep learning was calculated based on the likelihood of malignancy returned by the network. We used area under the receiver operating characteristic curve (AUC) as the performance metric. The calculation of AUC was achieved using pROC package in R. Confidence intervals were estimated using bootstrap [14] with 2000 stratified bootstrap replicates. Statistical comparison between the performances of radiologists and the performance of deep learning was conducted using DeLong Method [15]  In order to assess the impact of ultrasound scanner manufacturer and model, we repeated this analysis for three individual scanner types: ATL HDI 5000, GE LOGIQ E9, and Philips iU22 which together accounted for 95.5% of all images.

**Results**

*Comparison of deep learning and radiologist performance*

As shown in the second column of Table 2, using a new dataset, the AUC of our deep learning algorithm to distinguish benign from malignant nodules for all images combined was 0.69 (95% CI: 0.64-0.75). This performance was slightly higher than the AUC of radiologists, which ranged from 0.63-0.66 with an average of 0.64. The difference between performance of deep learning and average performance of radiologists was not statistically significant (p=0.40).

*Impact of Different Manufacturers for Deep Learning*

This new testing dataset varies significantly from the training dataset. In the training dataset, approximately 94% of the training data was from a Siemens scanner and less than 1% was from the Philips iU22 scanner which constituted significant majority of these test cases. Similarly, images from three main device types in the new testing dataset all constituted little in the training dataset. The performances of deep learning neural network and radiologists on images with three most common device types in the new testing dataset are shown in column 3 to column 5 in Table 2. For each of the US device types, the AUC of deep learning was comparable to the performance of individual radiologists, but slightly lower than the performance of radiologist average. and the differences between deep learning and average performance of radiologists were not significant (p = 0.32, 0.99, and 0.54 for three device types) for any of the scanners. We also observed some differences in the performance of both radiologists and deep learning for different scanners. For example, both radiologists average and deep learning achieved the highest AUC in images from LOGIQ E9 (0.79 and 0.75), and the lowest AUC in images from HDI 5000 (0.64 and 0.56) among the three main groups. Also, as shown in Table 3, based on the DeLong test [15] of radiologist response for three different scanners, the p-value of LOGIQ E9 vs HDI 5000 of both radiologist average and deep learning (0.08 and 0.03) is smaller than 0.1, indicating that though not statistically significant for the radiologist average, there is clearly a trend of significance of accuracy difference between these two different scanners for radiologist average and deep learning. For other scanner comparison, no significant difference is observed, though the comparison of deep learning between Philips iU22 and HDI 5000 also shows a substantial trend toward significance (p = 0.06).

*Agreement Between Radiologists*

Figure 3 shows the agreement between radiologists on thyroid nodule classification. We do see that some radiologists have minor disagreement between "Moderate chance of malignancy" and "Probably benign". However, the Cohen's kappa values between four radiologists are listed in Table 4, and all the kappa values are larger than 0.75, which shows that there is a moderate agreement between all the radiologists, though TI-RADS was not used in the study.

**Discussion**

Our study of 378 thyroid nodules on ultrasound aimed to evaluate the performance of a previously established deep learning model when using an entirely new dataset from a new institution. We found comparable performance between deep learning and radiologists. This was the case across all cases and individually for different scanners. Overall performance was slightly lower than expected for machine and humans alike.

In deep learning model development, using single institution data means that training and test sets have similar technical characteristics such as scanner type, imaging protocol, and procedures applied by technologists. These kinds of factors may differ at other institutions, and model performance may be significantly lower when applied in other environments. Furthermore, development of deep learning models is prone to overfitting to the data at hand, both in terms of the training of that model and algorithm design by humans. Therefore, some level of decrease in performance is expected. However, while our algorithm showed lower performance compared to what we previously reported, our radiologist readers also had relatively low performance for distinguishing benign from malignant nodules resulting in a similar performance between AI and radiologists as previously reported [6]. Based on excluding other likely reasons (see the discussion below), our belief is that the set of nodules considered in this study constituted a higher-than-average level of overall difficulty which contributed to decreased performance in both AI and radiologists.

Both the algorithm and radiologist readers had AUCs that differed across the three most common scanner types, and model types seemed to have similar effects for both algorithm and human alike. For example, both the neural network and radiologists had highest performance for scans from GE LOGIQ E9, followed by Philips iU22, and lowest performance from the ATL HDI 5000, as shown in Figure 4. These results suggest there may be inherent differences in imaging characteristics between different manufacturers, differences that might affect both a human eye as well as an algorithm. Indeed, there are perceptible differences in the appearance of images from each scanner, as shown in Figure 5 and 6.

Figure 5 has shown examples of thyroid ultrasound image input for the classification network. For each scanner, 5 nodules were randomly selected for comparison. The three columns in Figure 5 are from ATL HDI 5000, GE LOGIQ E9 and Philips iU22. As shown in Figure 5, all images after cropping and processing have the same number of pixels. Also, for comparison between before and after pre-processing, Figure 6 showed the original images of the three images in the first row of Figure 5 without any extra processing. From both Figure 5 and Figure 6, we can observe that there are obvious differences in calipers between different manufacturers even after pre-processing. However, the difference in calipers does not explain the different performance between scanners among radiologists since radiologists can easily identify the calipers from the images. We have also compared the performance of deep learning algorithm without and with manually contouring, and we found out that the performance is the same. It showed that the differences in calipers does not explain the different performance of the algorithm, either. Another possible reason is that images from ATL HDI 5000 has lower resolution than both GE LOGIQ E9 and Philips iU22. The lower resolution may affect the performance of radiologists for this group.

Please note that in order to increase the accuracy of the performance estimation, we used a set of nodules with a higher malignancy rate. The number of nodules used for model testing was 378 (756 ultrasound images), and 147 of them (39%) were malignant. This is a much higher percentage than that in both training dataset (9.9%) and testing dataset (15%) in the previous study. The percentage of malignant cases in the study was also much higher than the general malignancy percentage among thyroid nodules (4% - 6.5%) [15].

The differences of AUC between deep learning and average performance of radiologists were not significant ($p>0.3$) for any of the scanners. This demonstrates robustness of the algorithm on a variety of scanners despite the very different composition of the training set. We observed some difference in performance of both radiologists and deep learning for different scanners which may be caused by randomly more or less difficult sample or differences in quality of the images. For certain scanners, these differences have caused a considerable trend of significance on the performance for both radiologist average (p = 0.08) and deep learning (p = 0.04). The cause of these differences lie outside of the scope of our study.

**Conclusion**

The algorithm is generalized to images not seen in the training set. Even with the relatively difficult new dataset, the deep learning algorithm achieved similar AUC values for malignancy classification with all four radiologists. Though, certain manufacturer of ultrasound imaging shows a certain trend towards significant impact on the performance of both radiologists and the algorithm, it has not shown a significant impact on the relative performance difference between the algorithm and the radiologists.

**Tables**

| Image Characteristic | All Nodules | Benign Nodules | Malignant Nodules |
|---|---|---|---|
| Number of Nodules | 378 | 231 | 147 |
| Number of Nodules in Female Patients | 304 | 193 | 111 |
| Number of Nodules in Male Patients | 74 | 38 | 36 |
| Mean Age of Patients | 52.71 ± 14.71 | 54.38 ± 14.49 | 50.10 ± 14.65 |
| Maximum Nodule Size (cm) (± std. dev.) | 2.2 ± 1.2 | 2.3 ± 1.1 | 2.1 ± 1.2 |

**Table 1: Image Statistics**

| Image Groups/ Radiologists | All Images Combined (378 cases) | HDI 5000 (64 cases) | Philips iU22 (218 cases) | GE LOGIQ E9 (79 cases) |
|---|---|---|---|---|
| Radiologist 1 | 0.63 (0.59-0.67) | 0.58 (0.48-0.67) | 0.63 (0.57-0.69) | 0.68 (0.57-0.79) |
| Radiologist 2 | 0.66 (0.61-0.71) | 0.61 (0.48-0.74) | 0.67 (0.60-0.74) | 0.69 (0.58-0.80) |
| Radiologist 3 | 0.65 (0.60-0.70) | 0.60 (0.49-0.70) | 0.64 (0.58-0.70) | 0.70 (0.59-0.82) |
| Radiologist 4 | 0.63 (0.58-0.67) | 0.57 (0.48-0.66) | 0.63 (0.57-0.70) | 0.68 (0.59-0.77) |
| Radiologist Average | 0.72 (0.67-0.77) | 0.64 (0.50-0.77) | 0.72 (0.64-0.78) | 0.79 (0.68-0.90) |
| Deep Learning | 0.69 (0.64-0.75) | 0.56 (0.41-0.70) | 0.72 (0.64-0.79) | 0.75 (0.64-0.86) |
| p-value Between Radiologist Average and Deep Learning | 0.3994 | 0.3242 | 0.9883 | 0.5445 |

Table 2: AUCs of Radiologists and Deep Learning for All Images and Images from Three Main Device Types, p-value Indicates the Significance of AUC Difference between Radiologist Average and Deep Learning

(95% Confidence Interval in the Parenthesis)

| p value Between Different Scanners | LOGIQ E9 vs Philips iU22 | LOGIQ E9 vs HDI 5000 | Philips iU22 vs HDI 5000 |
|---|---|---|---|
| Radiologist 1 | 0.4190 | 0.1602 | 0.3576 |
| Radiologist 2 | 0.7544 | 0.3913 | 0.4712 |
| Radiologist 3 | 0.3819 | 0.2015 | 0.4855 |
| Radiologist 4 | 0.3852 | 0.0810 | 0.2597 |
| Radiologist Average | 0.2609 | 0.0820 | 0.3038 |
| Deep Learning | 0.5891 | 0.040 | 0.0587 |

Table 3: p-value Between Different Ultrasound Scanner of Radiologists and Deep Learning

| R1 vs R2 | R1 vs R3 | R1 vs R4 | R2 vs R3 | R2 vs R4 | R3 vs R4 |
|---|---|---|---|---|---|
| 0.7593 | 0.8730 | 0.8228 | 0.8095 | 0.7857 | 0.8413 |

Table 4: Cohen's kappa Values of Radiologists

(R1,2,3,4 denotes four radiologists)

**Figures**

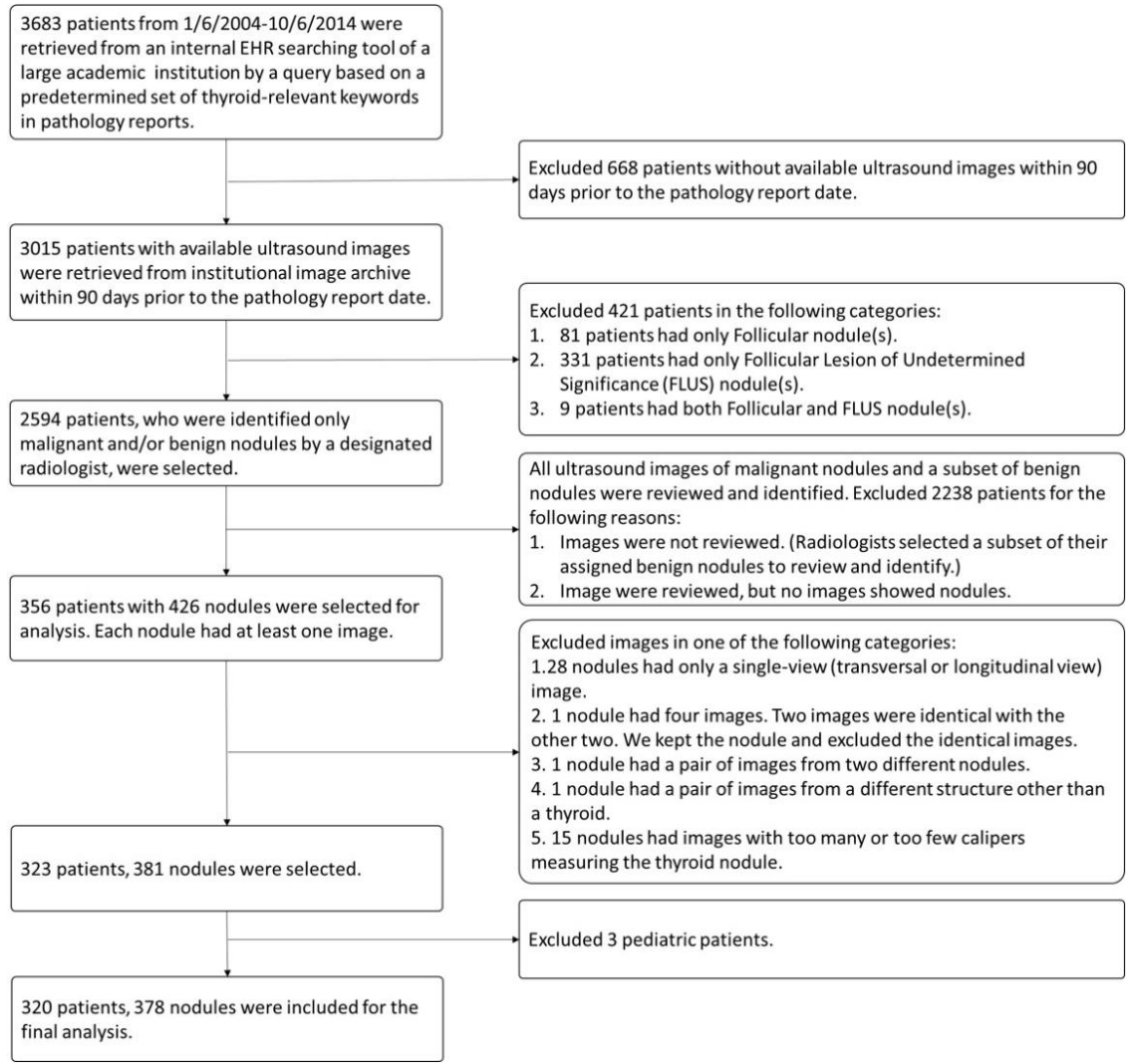

Fig. 1: Image Exclusion Flowchart

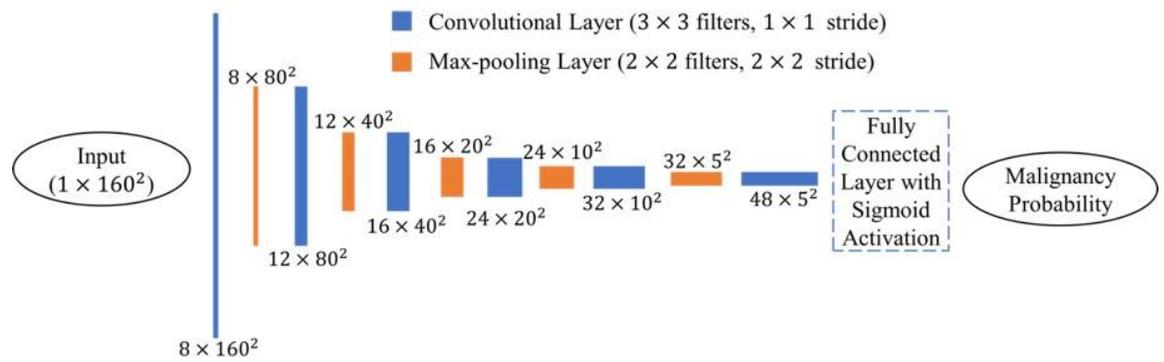

Fig. 2: Classification Network Architecture

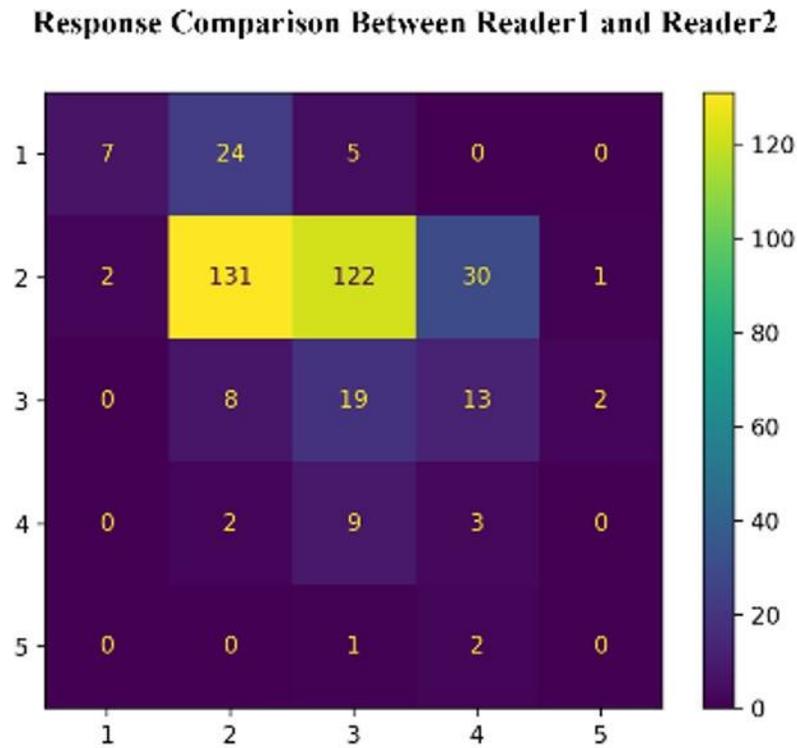

Fig. 3A: Response Comparison between Reader 1 and Reader 2

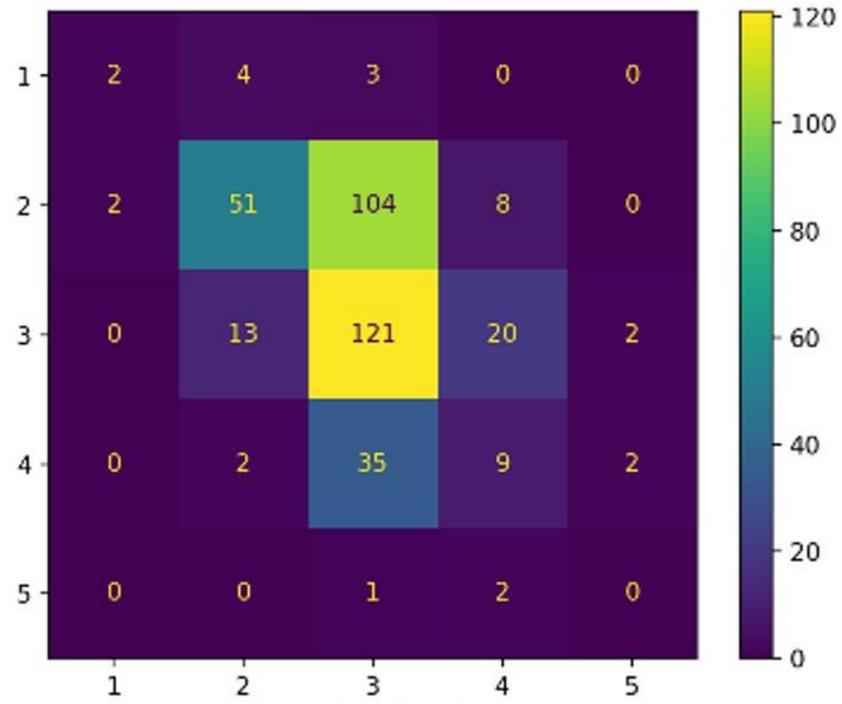

Fig. 3B: Response Comparison between Reader 1 and Reader 3

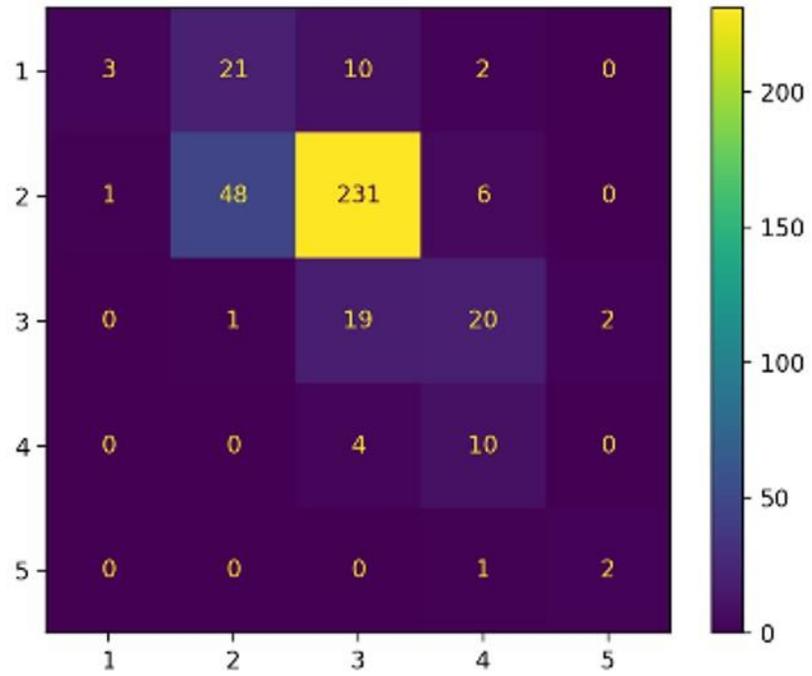

Fig. 3C: Response Comparison between Reader 1 and Reader 4

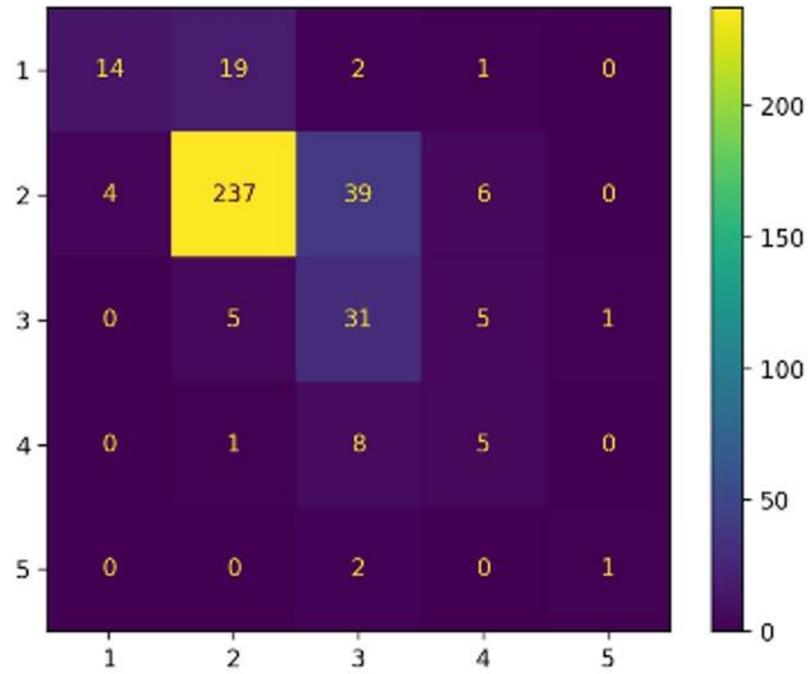

Fig. 3D: Response Comparison between Reader 2 and Reader 3

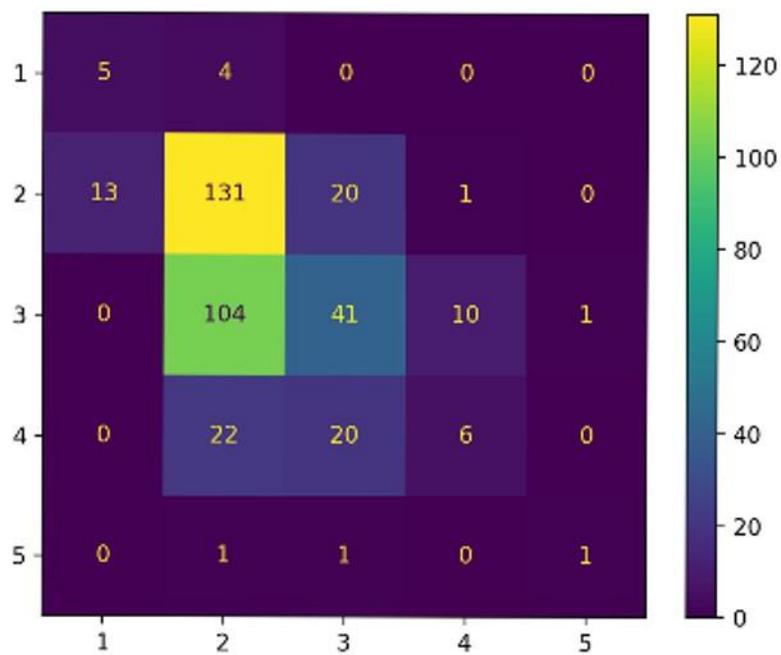

Fig. 3E: Response Comparison between Reader 2 and Reader 4

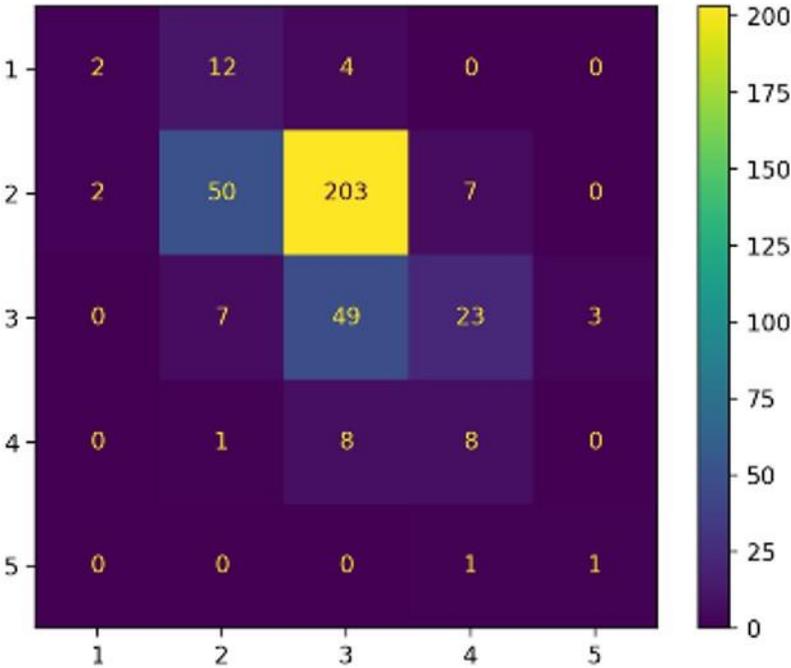

Fig. 3F: Response Comparison between Reader 3 and Reader 4

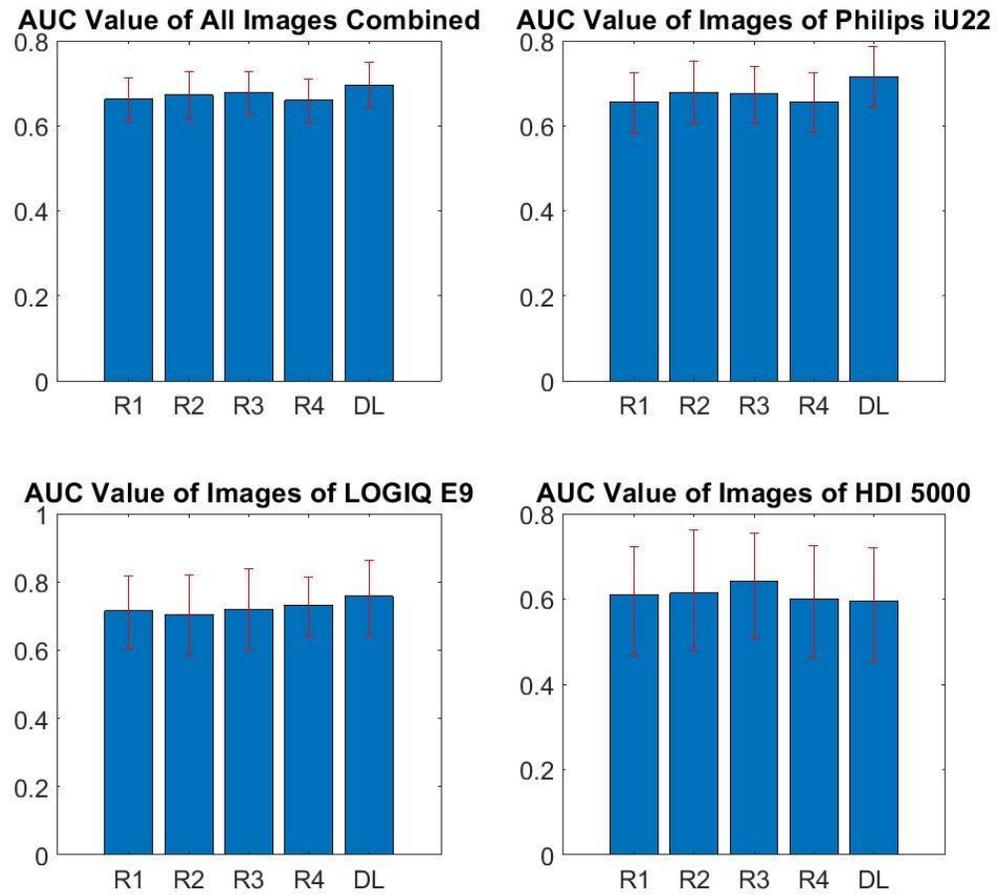

Fig. 4: AUC (Area under curve) performance of radiologists and deep learning for the entire validation set and cases acquired using specific US scanners

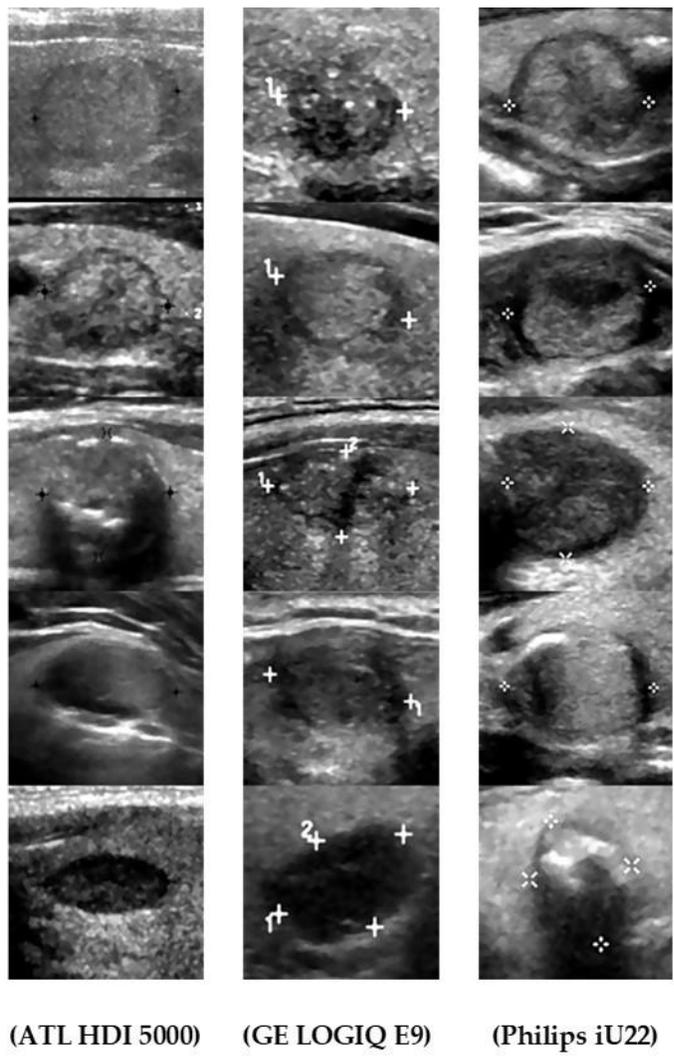

Fig. 5: Comparison of Cropped Ultrasound Images from Three Manufacturers with Pre-processing

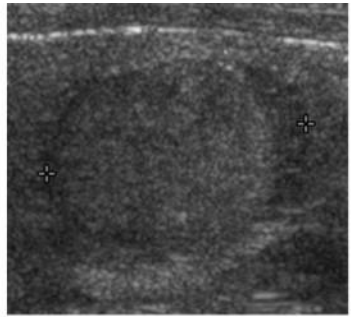 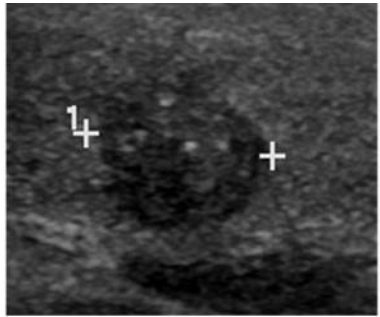 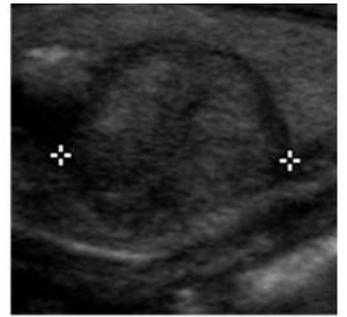

(ATL HDI 5000)　　　　　(GE LOGIQ E9)　　　　　(Philips iU22)

Fig. 6：Ultrasound Images from Three Manufacturers Without Pre-processing